\documentclass[amsmath,amssymb,aps,prb,citeautoscript,superscriptaddress,twocolumn,showpacs,10pt]{revtex4-1}

\usepackage{bm}
\usepackage{graphicx}
\usepackage{dcolumn}
\usepackage[colorlinks,linkcolor=red,anchorcolor=green,
citecolor=blue,breaklinks]{hyperref}

\usepackage{mathrsfs}
\usepackage{color}

\graphicspath{{figures/}}
\begin{document}

\preprint{VPT}

\title{Dissipative quantum transport at arbitrary parameter regime: a variational method}

\author{Yu Zhang}
\email{zhy@yangtze.hku.hk}
\altaffiliation[Present institute: ]{Center of Bio-inspired Energy Science, Northwestern University, Evanston, IL, USA.}
\affiliation{
Department of Chemistry, The University of Hong Kong,
Pokfluam Road, Hong Kong
}

\author{ChiYung Yam}
\affiliation{
Beijing Computational Science Research Center,
No. 3 He-Qing Road, Beijing 100084, China}%
\affiliation{
Department of Chemistry, The University of Hong Kong,
Pokfluam Road, Hong Kong
}

\author{YanHo Kwok}
\affiliation{
Department of Chemistry, The University of Hong Kong,
Pokfluam Road, Hong Kong
}

\author{GuanHua Chen}
\email{ghc@everest.hku.hk}
\affiliation{
Department of Chemistry, The University of Hong Kong,
Pokfluam Road, Hong Kong
}

\date{\today}

\begin{abstract}
Recent development of theoretical method for dissipative
quantum transport have achieved notable progresses
in the weak or strong electron-phonon coupling regime.
However, a generalized theory for dissipative quantum
transport at arbitrary parameter regime is not figured out
until now. In this work, a variational method for dissipative
quantum transport at arbitrary electron-phonon
coupling regime is developed by employing variational
polaron theory. The optimal polaron transformation is
determined by the optimization of the Feynman-Bogoliubov
upper bound of free energy.
The free energy minimization ends up with an optimal mean-field
Hamiltonian and a minimal interaction Hamiltonian. Hence,
second-order perturbation can be applied to the transformed
system, resulting an accurate and efficient method for the
treatment of dissipative quantum transport with arbitrary
electron-phonon coupling strength. Numerical benchmark
calculation on an single site model with coupling to one
phonon mode is also presented.
\end{abstract}


\maketitle

\section{Introduction}
\label{seq:intro}
Quantum transport and energy dissipation in nanostructures
have been of great interest, which requires the study
and understanding of nonequilibrium phenomena of open
quantum systems~\cite{kumar2012prl,agra2002prl,magnus2008prl,
Dubirmp2011,roman02010prb,jcp164121}.
In many previous studies on dissipative quantum transport,
the coupling between the system and the phonon bath
can be small. In this case, second-order perturbation (2PT)
can be employed, leading to a
set of equation of motions (EOM)~\cite{jcp164121}.
However, there are many open quantum
systems with strong electron-phonon coupling. For instance,
the coupling between system and bath in the energy transfer
process in photosynthetic complexes is comparable to the
electronic coupling~\cite{brixner2005,njp105012,jp050788d};
Strong electron-phonon coupling is also observed in resonant
tunneling in molecular transport junctions~\cite{nl048216c,
parknature2000}.
Even though there are several non-perturbative methods to
evaluate the dynamics numerically exactly, such as hierarchical
equations of motion (HEOM)~\cite{JPSJ.75.082001,JPSJ.74.3131},
quasi-adiabatic propagator path integral
(QUAPI)~\cite{1.469508,*1.469509} and multiconfiguration
time-dependent Hartree (MCTDH) approach~\cite{Beck20001,
thoss2001jcp,Meyer199073}, these methods are computational
demanding and not trivial to be implemented.

There is another group of methods to deal with strong
electron-phonon coupling, which makes use of polaron
transformation. A polaron
transformed second-order master equation for open quantum system
has been developed accordingly~\cite{PhysRevB.83.165101,
*nazir113042,PhysRevLett.103.146404,1.3247899,
*1.3608914,1.2977974}. The polaron theory has also been
employed to study the quantum transport in the strong
coupling regime~\cite{PhysRevB.40.11834,PhysRevB.66.075303,
PhysRevB.71.165324,PhysRevB.67.165326,nitzan2006prb,
PhysRevB.67.165326,PhysRevB.77.205314,*PhysRevB.87.085422,
zhangpolaron}, in which strong phonon blockade effect is
observed. The polaron transformation extends the validity
of perturbative
treatment of electron-phonon interaction to strong
coupling regime, provided that the electronic couplings are
small compared to the typical electron-phonon coupling and
bath frequency. However, when this condition is not fulfilled,
the polaron theory performs worse than the standard perturbative
approaches as the polaron cannot follow the system
motion accurately.

In order to overcome this shortcoming of polaron
transformation based approaches, the variational
method has been developed as a generalized polaron
transformation~\cite{PhysRevB.84.081305,nazir075018,
1.3636081,1.447055,*1.449469,1.4722336}. In contrast with
full polaron transformation, the variational polaron approach
searches for an optimal amount of transformation by employing
variational principle, which is implemented by minimization
of the Feynman-Bogolubov upper bound of free energy~\cite{1.447055,*1.449469}.
The optimal transformation is determined by the properties of the bath
and system-bath coupling. Thanks to the variational principle,
the variational polaron method is able to interpolate
between the weak and strong coupling regimes and to capture
the dynamic behaviour at arbitrary parameter space.
Moreover, compared to the many-body approaches, variational
polaron method is computationally economic.

Inspired by the recently developed variational polaron master
equation~\cite{PhysRevB.84.081305,nazir075018},
variational polaron method for dissipative quantum transport
is developed in this work, which aims at handling the
dissipative quantum transport at arbitrary electron-phonon
coupling strength. The manuscript is organized as follows.
Sec.~\ref{sec:vpt} introduces the variational polaron
theory for quantum transport and minimization method for
determining the optimal parameters for variational
transformation. Starting from the variational polaron
transformation, the transformed Hamiltonian is
divided into two parts: a mean-field part and
interacting part. Free energy optimization ends up
with an optimal mean-field Hamiltonian and a small interacting
perturbation. Then 2PT is applied
to establish the quantum transport theory.
The variational polaron theory for quantum transport is
applied to study the quantum transport through a single
site model with different electron-phonon coupling strength,
which is given in Sec.\ref{sec:numerical}.
Finally, Sec.\ref{sec:summary} summarizes the method
developed in this work.

%
\section{Variational polaron theory for quantum transport}
\label{sec:vpt}
\subsection{Model Hamiltonian}
Considering a system sandwiched between two leads. The
electrons can transfer from one lead to the other through
the system driven by bias voltage. The electrons
can be scattered inelastically by phonons when transport
through the system.
The corresponding Hamiltonian reads
\begin{equation}
H=H_S + \sum_\alpha[H_\alpha+H_{S\alpha}] + H_{B}+H_{SB},
\end{equation}
where $H_S$ is the Hamiltonian of the system; $H_{B}$ is the
Hamiltonian of phonon bath; $H_\alpha$ is the Hamiltonian of
lead $\alpha$. The lead serves as electronic bath; $H_{SB}$
and $H_{S\alpha}$ are the Hamiltonians of system-phonon
coupling and system-lead interaction, respectively.
Expressions of those Hamiltonians are
\begin{eqnarray}\label{eq:ham}
H_S=&&\sum_n \epsilon_n c^\dag_n c_n, \nonumber\\
H_\alpha=&& \sum_{k_\alpha}\epsilon_{k_\alpha}c^\dag_{k_\alpha} c_{k\alpha}
\nonumber\\
H_{S\alpha}=&&\sum_{n,k_\alpha}[V_{k_\alpha n}c^\dag_{k_\alpha}c_n+\text{H.c.}],
\nonumber\\
H_B=&&\sum_{n,k} \omega_{n,k} b^\dag_{n,k} b_{n,k}
\nonumber\\
H_{SB}=&&\sum_{n,k}g_{n,k} c^\dag_n c_n(b^\dag_{n,k} +b_{n,k}).
\end{eqnarray}
Where $\epsilon_n$ denotes the single-particle energy of the
electronic state $n$ in the device and $c^\dag_n$ ($c_n$)
is the corresponding creation (annihilation) operator.
Similarly, $\epsilon_{k_\alpha}$ and $c^\dag_{k_\alpha}$
($c_{k_\alpha}$) represent the single-particle energy and
creation (annihilation) operator of $k$th electronic state
in lead $\alpha$, respectively. $V_{k_\alpha n}$ is
the system-lead coupling strength. The phonon creation and
annihilation operators are represented by $b^\dag_{n,k}$ and
$b_{n,k}$. And
$\omega_{n,k}$ and $g_{n,k}$ are the corresponding the phonon
energy and electron-phonon coupling constant.

In the absence of electron-phonon coupling, an exact
time-dependent quantum transport theory has been developed
recently~\cite{zheng2010jcp,PhysRevB.75.195127,C1CP20777F,
PhysRevB.87.085110,4737864}. 
In presence of electron-phonon
coupling, electron has the probability of being scattered
inelastically by phonon. Previous theoretical studies of
dissipative quantum transport usually focus on two different
regimes: weak or strong electron-phonon coupling. In the weak
coupling regime, 2PT (or lowest order expansion) is widely
employed. While in the strong electron-phonon coupling limit,
the electron is localized by phonon scattering. At this regime,
polaron theory is suitable for describing the phenomena.
In this manuscript, a generalized theory for dissipative
quantum transport with arbitrary electron-phonon coupling
is developed by extending the polaron theory to its
variational version.

\subsection{Variational polaron transformation}
Similar to the polaron theory, the variational polaron theory
starts from an unitary transformation generated by
the operator
\begin{equation}
U=\text{exp}\left[
\sum_{n,k}\lambda_{n,k}c^\dag_nc_n(b^\dag_{n,k}-b_{n,k})
\right],
\end{equation}
where $\lambda_{n,k}=f_{n,k}/\omega_{n,k}$.
The unitary operator displaces the phonon oscillation in the
positive and negative direction. The parameter $f_{n,k}$
determines the magnitude of the displacement for each mode.
It is obvious that $f_{nk}=g_{nk}$ corresponds to the
conventional polaron transformation, while $f_{nk}=0$
denotes no transformation.

After the unitary transformation, the Hamiltonian of the
system becomes
\begin{equation}
\bar{H}_S=\sum_n (\epsilon_n+R_n) c^\dag_n c_n,
\end{equation}
where $R_n=\lambda_{n,k}(f_{n,k}-2g_{n,k})$ is the
renormalization energy which describes the shift of
single-particle energy of electronic states induced by
electron-phonon interaction. $\bar{H}_\alpha$
is same as $H_\alpha$ since it commutes with the unitary
operator $U$. While, the system-lead coupling Hamiltonian
becomes
\begin{equation}
\bar{H}_{S\alpha}=\sum_{nk_\alpha}[V_{k_\alpha n}
c^\dag_{k_\alpha}c_n X_n +\text{H.c.}]
\end{equation}
after the transformation, i.e., the system-lead
coupling is dressed by phonon displacement operator $X_n$,
where $X_n$ is defined as
\begin{equation}
X_n=\text{exp}\left[
-\sum_k \lambda_{n,k}(b^\dag_{n,k}-b_{n,k})
\right].
\end{equation}
After the unitary transformation, the Hamiltonian of
electron-phonon coupling becomes
\begin{equation}
\bar{H}_{SB}=\sum_{n,k}\bar{g}_{n,k}c^\dag_nc_n
(b^\dag_{n,k}+b_{n,k}),
\end{equation}
where $\bar{g}_{n,k}=g_{n,k}-f_{n,k}$ is defined as
the reduced electron-phonon coupling strength.
$\bar{g}_{n,k}$ indicates that the electron-phonon
coupling is reduced by the unitary transformation. In summary,
after the variational polaron transformation, phonon's
influence on electrons is transferred into three terms:
(1) the polaron shift energy $R_n$, which represents the energy
renormalization effect due to electron-phonon interaction;
(2) in the dressed coupling between system and leads, the
system-lead coupling is renormalized by the displacement
operator $X_n$;
(3) reduced electron-phonon coupling $\bar{g}_{n,k}$,
which is zero in the strong coupling limit and approaches
$g_{nk}$ in the weak coupling limit.

The variational transformation results in a new kind of
system-lead coupling, which is mediated by the phonon
displacement operator. The displacement operator can be
divided into two parts:
\[
X_n=X_n-\langle X_n\rangle + \langle X_n\rangle
\equiv B_n + \langle X_n\rangle,
\]
i.e., the expectation value and
corresponding fluctuation around its expectation.
Assuming phonon is in thermal equilibrium,
$\langle X_n\rangle$ can be written as
\begin{equation}
\langle X_n\rangle =\text{exp}\left[
-\frac{1}{2}\sum_k \lambda^2_{n,k}
\text{coth}(\beta\omega_{n,k}/2)\right],
\end{equation}
where $\beta$ is the inverse temperature. Consequently,
$H_{S\alpha}$ can be separated into two parts,
\begin{eqnarray}\label{eq:ham2}
\bar{H}^0_{S\alpha}=&&\sum_{n,k_\alpha}[V^R_{k_\alpha n}
c^\dag_{k_\alpha}c_n+\text{H.c.}],
\nonumber\\
\bar{H}^I_{S\alpha}=&&\sum_{n,k_\alpha}[V_{k_\alpha n}
c^\dag_{k_\alpha}c_n B_n+\text{H.c.}].
\end{eqnarray}
where $V^R_{k_\alpha n}=V_{k_\alpha n}\langle X_n\rangle$
is the renormalized system-lead coupling mediated by phonon.
Thus, the total transformed Hamiltonian can also be grouped
into two parts. One part is the mean-field (or non-interacting)
part $\bar{H}_0$ and the other one is the interacting
Hamiltonian $\bar{H}_I$, where $\bar{H}_0$ is defined as
\begin{equation}
\bar{H}_0=\bar{H}_S  + \sum_\alpha(\bar{H}^0_{S\alpha}+\bar{H}_\alpha)
+ \bar{H}_B\equiv \bar{H}_e+\bar{H}_B.
\end{equation}
and $\bar{H}_I$ is
\begin{equation}
\bar{H}_I=\bar{H}_{SB}+\sum_\alpha \bar{H}^I_{S\alpha}.
\end{equation}

It is obvious that the polaron transformation
($f_{n,k}=g_{n,k}$) makes the electron-phonon
interaction term $\bar{H}_{SB}$ vanished. In contrast,
the variational polaron transformation employs $f_{nk}$ as
parameters and the values of which are determined by the
minimization of an upper bound of free energy. The minimization
inherent to variational approach allows to minimize the
effect of $\bar{H}_I$ and obtain an optimal mean-field
Hamiltonian $\bar{H}_0$. Thus, given
arbitrary electron-phonon coupling strength, the variational
polaron transformation in principle can end up with small
$\bar{H}_I$. Consequently, 2PT can
be employed to account the effect of $\bar{H}_I$.
Because of the single-particle nature of $\bar{H}_0$ and
second-order perturbative treatment of $\bar{H}_I$,
variational polaron theory significantly reduces the
computational complexity compared to the many-particle
approaches.

\subsection{Minimization of the upper bound of free energy}
In this work, the Feynman-Bogoliubov upper bound of free
energy~\cite{1.447055,*1.449469}
is chosen to determine the optimal value of $f_{nk}$
for the variational transformation.
The upper bound of free energy is given as
\begin{equation}\label{eq:free1}
F_u=-\frac{1}{\beta}\text{ln}\{\text{tr}[e^{-\beta \bar{H}_0}]\}
+\langle \bar{H}_I \rangle_{\bar{H}_0},
\end{equation}
where $\langle \bar{H}_I \rangle_{\bar{H}_0}=\text{Tr}[
\bar{H}_I e^{-\beta\bar{H}_0}]$. $F_u$ gives the upper bound
of the true Free energy $F$, which is related to $F$ by the
inequality $F\leq F_u$. It is obvious that the second term
on the right hand right (RHS) of
Eq.(\ref{eq:free1}) is zero by the construction.
Besides, because $[\bar{H}_B,\bar{H}_e]=0$, $F_u$ can be
further simplified and divided into two parts
\begin{equation}\label{eq:free2}
F_u=F_{\text{B}}-\frac{1}{\beta}\text{ln}\{\text{tr}
[e^{-\beta \bar{H}_e}]\},
\end{equation}
where $F_B$ is the free energy of phonon bath. Because $F_B$
is not a function of $f_{nk}$, it is neglected in the
minimization procedure. There are two ways to minimize
the upper bound of free energy with respect to $f_{nk}$.
One way is to treat the system and leads as a whole system
and the upper bound of free energy can be numerically
calculated by the diagonalization of $\bar{H}_e$. The other
approach is to employ the Linked cluster expansion
method~\cite{mahan2000book}.

%
The mean-field part of the transformation Hamiltonian,
$\bar{H}_0$,
is a function of the phonon-induced renormalization parameters
$\{R_n, \langle X_n\rangle\}$. Consequently, the
Feynman-Bogoliubov upper bound
of free energy $F_u$ can be minimized with respect to
$\{R_n, \langle X_n\rangle\}$. Hence, the minimization
condition can be written as
\begin{equation}\label{eq:freeminimization}
\frac{dA_S}{df_{n,k}}=\frac{\partial A_S}{\partial R_n}
\frac{\partial R_n}{\partial f_{n,k}}+
\frac{\partial A_S}{\partial \langle X_n\rangle}
\frac{\partial \langle X_n\rangle}{\partial f_{n,k}}=0.
\end{equation}
Using the expressions for $\{R_n, \langle X_n\rangle\}$, we can
 evaluate the derivatives of $\{R_n, \langle X_n\rangle\}$
with respect to $f_{n,k}$ analytically.
After simple algebra, the minimization condition gives the
expression of the variational transformation parameter
$f_{n,k}$ as $f_{n,k}=g_{n,k}F_{n,k}$,
where
\begin{equation}
F_{n,k}=\frac{\omega_{n,k}\frac{\partial A_S}{\partial R_n}}
{\omega_{n,k}\frac{\partial A_S}{\partial R_n}-
\frac{1}{2}\frac{\partial A_S}{\partial \langle X_n\rangle}
\text{coth}(\beta\omega_{n,k}/2)\langle X_n\rangle}.
\end{equation}
Since $\{R_n,\langle X_n\rangle\}$ are also functions
of $f_{nk}$, above equation must be solved self-consistently.
After the minimization procedure, the optimal choice of
$\bar{H}_0$ is obtained, resulting in a small $\bar{H}_I$.
Hence, $\bar{H}_I$ can be treated by 2PT.
Next, dissipative quantum transport formalism with variational
polaron transformation is presented.

%
\section{Dissipative quantum transport theory with variational
polaron transformation}
\label{sec3}
Non-equilibrium Green's function (NEGF) formalism
is employed to study the dissipative quantum transport.
The key quantity in the NEGF method is the
single-particle Green's function, which is given by
\begin{eqnarray}\label{eq:greenfunctionc}
G_{ij}(\tau,\tau')=&&-i\langle T_c c_i(\tau)c^\dag_j(\tau')\rangle_H
\nonumber\\
=&&-i\langle T_c c_i(\tau)X_i(\tau)c^\dag_j(\tau')X^\dag_j(\tau')\rangle_{\bar{H}},
\end{eqnarray}
where $H$ and $\bar{H}$ are the Hamiltonians before and after
the variational polaron transformation, respectively.
$\tau$ and $\tau'$ are the time variables defined on the
Keldysh contour and $T_c$ is the contour time-ordering
operator. Eq.(\ref{eq:greenfunctionc}) determines the dynamics
of the coupled electrons and phonons. We employ the following
approximation to decouple the electron and phonon dynamics
\begin{equation}\label{eq:green0}
G_{ij}(\tau,\tau')=\bar{G}_{ij}(\tau,\tau')K_{ij}(\tau,\tau'),
\end{equation}
where
\begin{eqnarray}
\bar{G}_{ij}(\tau,\tau')=&&-i\langle T_c c_i(\tau)c^\dag_j(\tau')\rangle_{\bar{H}}
\nonumber\\
K_{ij}(\tau,\tau')=&& \langle T_c X_i(\tau)X^\dag_j(\tau')\rangle_{\bar{H}}.
\end{eqnarray}
The decoupling in Eq.(\ref{eq:green0}) is inherent in the Born-Oppenheimer
approximation. Even the decoupling approximation is made, there is still
correlation between electron and phonon if self-consistent procedure is
operated \cite{nitzan2006prb}, which is similar to the diagram dressing
process in the standard many-body perturbation theory.

With the decoupling approximation, the key quantity becomes
$\bar{G}(\tau,\tau')$ which is associated with $\bar{H}$.
As illustrated in Sec.~\ref{sec:vpt}, the variational polaron
transformed Hamiltonian can be divided into two parts:
$\bar{H}_0$ and $\bar{H}_I$. $\bar{H}_0$ is mean-field
part of the Hamiltonian, the quantum
transport problem associated with $\bar{H}_0$ can be easily
solved as demonstrated by previous publications~\cite{PhysRevB.50.5528}.
While the correction of $\bar{H}_I$ to the quantum transport
can be taken into consideration by the perturbation theory
since the variational polaron transformation in principle
ends up with small $\bar{H}_I$. Firstly, zero-order Green's
function  $G_0(\tau,\tau')$ is obtained by using $\bar{H}_0$,
\begin{eqnarray}
G_0(\tau,\tau')=&&g_0(\tau,\tau')+\int d\tau_1\int d\tau_2
g_0(\tau,\tau_1)
\nonumber\\ &&\times
\left[\sum_\alpha\bar{\Sigma}_{0\alpha}(\tau_1,\tau_2)\right]
G_0(\tau_2,\tau'),
\end{eqnarray}
where $g_0(\tau,\tau')$ is the electron Green's function
in absence of system-lead coupling. The effect of system-lead
coupling is represented by the self-energy
\begin{eqnarray}\label{eq:self0}
\bar{\Sigma}_{0\alpha,ij}(\tau,\tau')=&&\sum_{k_\alpha}
V^*_{k_\alpha i}V_{k_\alpha j}g_{k_\alpha}(\tau,\tau')
\langle X_i\rangle \langle X_j\rangle
\nonumber\\=&&
\Sigma_{\alpha,ij}(\tau,\tau')
\langle X_i\rangle \langle X_j\rangle,
\end{eqnarray}
where $g_{k_\alpha}(\tau,\tau')$ is the surface Green's
function of lead $\alpha$ and $\Sigma_\alpha(\tau,\tau')$
is the lead self-energy in absence of phonon renormalization.
Next, $G(\tau,\tau')$ is derived at second-order expansion
with respect to $\bar{H}_I$, starting from $G_0(\tau,\tau')$.

%
\subsection{Second-order perturbation theory}
Within the 2PT theory, the Green's
function can be expressed as
\begin{eqnarray}
\bar{G}(\tau,\tau')=\bar{G}_0(\tau,\tau')+&&
\int d\tau_1\int d\tau_2 \bar{G}_0(\tau,\tau_1)
\nonumber\\&&\times
\bar{\Sigma}_I(\tau_1,\tau_2)\bar{G}_0(\tau_2,\tau'),
\end{eqnarray}
where $\bar{\Sigma}_I(\tau,\tau')$ is the self-energy to Green's
function at the second-order with respect to $\bar{H}_I$ and
$\bar{G}_0(\tau,\tau')$ is the Green's function corresponding
to $\bar{H}_0$. Since $\bar{H}_I$ contains two parts,
$\bar{H}^I_{S\alpha}$ and $\bar{H}_{SB}$,
$\bar{\Sigma}_I(\tau,\tau')$ can also be divided into two
parts accordingly,
\[
\bar{\Sigma}_I(\tau,\tau')=
\bar{\Sigma}_{ep}(\tau,\tau')+\sum_\alpha
\bar{\Sigma}_{I\alpha}(\tau,\tau').
\]
Within the 2PT,
$\Sigma_{I}(\tau,\tau')$ is derived as
\begin{eqnarray}
\bar{\Sigma}_{I\alpha,ij}(\tau,\tau')=&&\sum_{k_\alpha}
V^*_{k_\alpha i}V_{k_\alpha j}g_{k_\alpha}(\tau,\tau')
\nonumber\\&&\times
\langle T_C B_j(\tau')B^\dag_i(\tau)\rangle_{\bar{H}_0}.
\end{eqnarray}
The definition of $\bar{\Sigma}_{I\alpha}$ is very similar
to Eq.(\ref{eq:self0}) except the phonon renormalization
factor $\langle X_i\rangle$ is replaced
by the correlation function of $B_i$.
According to the definition of $B_i$,
the self-energy $\Sigma_{\alpha,I}$ can be rewritten as
\begin{eqnarray}
\bar{\Sigma}_{I\alpha,ij}(\tau,\tau')=&&
\Sigma_{\alpha,ij}(\tau,\tau')[K_{ji}(\tau',\tau)-
\langle X_i\rangle\langle X_j\rangle]\nonumber\\
=&&\Sigma_{\alpha,ij}(\tau,\tau')K_{ji}(\tau',\tau)-
\bar{\Sigma}_{0\alpha,ij}(\tau,\tau'),
\end{eqnarray}
where $\bar{\Sigma}_{0\alpha}(\tau,\tau')$ is the phonon
dressed self-energy, which is used to evaluate the zero-order
Green's function $\bar{G}_0(\tau,\tau')$ as described
previously.

The derivation of $\bar{\Sigma}_{ep}(\tau,\tau)$ is
same as that in Ref.~\onlinecite{jcp164121}. The only
difference is that the electron-phonon coupling
constant ($g_{n,k}$) is replaced by the reduced one
($\bar{g}_{n,k}$). The expression of
$\bar{\Sigma}_{ep}(\tau,\tau')$ is
\begin{equation}
\Sigma_{ep}(\tau,\tau')=
\sum_{n,k}\bar{g}_{n,k} D_0(\tau,\tau')
\bar{G}_0(\tau,\tau')\bar{g}_{n,k},
\end{equation}
where $D_0(\tau,\tau')$ is the bare phonon Green's function.

In practical implementation, the Green's function and
self-energies defined on the Keldysh contour are projected
on real axis, resulting in equations for retarded/advanced
and lesser/greater Green's functions and self-energies.
In this work, we focuses on the steady state. In steady
state, all quantities (two-time functions) are only
dependent on the time difference instead of two times.
Hence, Fourier transformation is employed to transfer
the quantities from real-time domain to energy space.
Since $\bar{H}_0$ is the mean-field part of the total
Hamiltonian, the corresponding Green's functions can be
easily evaluated,
\begin{eqnarray}
\bar{G}^r_0(E)=&&[E+i\eta-\bar{H}_0-\sum_\alpha
\bar{\Sigma}^r_{0\alpha}(E)]^{-1},
\nonumber\\
\bar{G}^<_0(E)=&&\bar{G}^r_0(E)[\sum_\alpha
\bar{\Sigma}^<_{0\alpha}(E)]\bar{G}^a_0(E),
\end{eqnarray}
where $\eta\rightarrow 0^+$.
$\bar{\Sigma}^r_{0\alpha}$ and $\bar{\Sigma}^<_{0\alpha}$
are phonon dressed retarded and lesser self-energies of lead
$\alpha$, respectively, which are written as
\begin{eqnarray}
\bar{\Sigma}^{r/<}_{\alpha,ij}(E)=&&\sum_{k_\alpha}
V^*_{ik_\alpha}V_{k_\alpha j}g^{r/<}_{k_\alpha}(E)
\langle X_i\rangle \langle X_j\rangle
\nonumber\\=&&
\Sigma^r_{\alpha,ij}(E)\langle X_i\rangle\langle X_j\rangle
\end{eqnarray}
Defining the unperturbated line-width function as
$\Gamma_\alpha(E)=-2\Im[\Sigma^r_\alpha(E)]$, where
$\Im$ denotes the imaginary part. Thus, the phonon dressed
line-width function becomes $\bar{\Gamma}_{\alpha,ij}(E)=
\Gamma_{\alpha,ij}(E)\langle X_i\rangle\langle X_j\rangle$.
The dressing factor indicates the localization of electron
induced by the electron-phonon interaction. In term of
line-width function, the corresponding lesser self-energy
can be rewritten as
$
\bar{\Sigma}^<_\alpha(E)=if_\alpha(E)\bar{\Gamma}_\alpha(E)
$
according to the fluctuation-dissipation law,
where $f_\alpha(E)$ is the Fermi-Dirac distribution function
of lead $\alpha$ with chemical potential $\mu_\alpha$.

After obtaining the zeroth-order Green's functions,
$\bar{G}^{r/<}_0(E)$, are
evaluated, the Green's function with $\bar{H}_I$ is obtained
from the 2PT theory as
\begin{eqnarray}
\bar{G}^r(E)=&&\bar{G}^r_0(E)+\bar{G}^r_0(E)
\bar{\Sigma}^r_I(E)\bar{G}^r_0(E),
\nonumber\\
\bar{G}^<(E)=&&\bar{G}^<_0(E)+\bar{G}^r_0(E)
\bar{\Sigma}^<_I(E)\bar{G}^<_0(E).
\end{eqnarray}
where $\bar{\Sigma}^{r/<}_I(E)$ is the self-energy to Green's
function at the second-order with respect to $\bar{H}_I$.
As shown previously, $\bar{\Sigma}^{r/<}_I(E)$ contains two
parts,
\[
\bar{\Sigma}^{r/<}_I(E)=\bar{\Sigma}^{r/<}_{ep}(E)+
\sum_\alpha \bar{\Sigma}^{r/<}_{I\alpha}(E).
\]
Retarded and lesser phonon self-energies within the
2PT are~\cite{jcp164121,jz5003154,4825226}
\begin{eqnarray}
\bar{\Sigma}^<_{ep}(E)=&&\sum_{q,\pm}\bar{g}_q N^\pm_q
\bar{G}^<_0(E\pm\omega_q) \bar{g}_q,\nonumber\\
\bar{\Sigma}^r_{ep}(E)=&&\sum_{q,\pm}\bar{g}_q[N^\mp_q
\bar{G}^r_0(E\pm\omega_q)\pm \bar{G}^<_0(E\mp\omega_q)]\bar{g}_q,
\end{eqnarray}
where $N^\pm_q = N_q+\frac{1}{2}\pm\frac{1}{2}$.
$N_q$ is the phonon occupation number of mode $q$,
which is determined by Bose-Einstein distribution function.

The derivation of $\Sigma^{r/<}_{I\alpha}(E)$ requires the
knowledge of the displacement correlation function
$K(\tau,\tau')$, which is nontrivial.
$K(\tau,\tau')$ in principle
depends on the phonon Green's function which is coupled with
electronic Green's function via its
self-energy~\cite{nitzan2006prb,PhysRevB.77.205314,*PhysRevB.87.085422,zhangpolaron}.
Therefore, self-consistent calculation of phonon and
electron Green's function is required. However numerical
implementation of self-consistent calculation is
non-trivial and very computational demanding. In practise,
we assume the phonon is in equilibrium and undressed by the
electron in this work. The influence of electron to the
phonon can be introduced through a phenomenological rate
equation including the renormalization, damping and heating
effect~\cite{PhysRevB.72.201101,PhysRevLett.93.256601,
Kristen2013}. With the assumption that phonon is in the
equilibrium and undressed by electron, the displacement
correlation function can be rewritten in a simple form~\cite{mahan2000book,
PhysRevB.71.165324,nitzan2006prb}. The lesser projection of
displacement correlation function is expressed as
\begin{eqnarray}\label{eq:shiftcorrelation1}
K^<_{ij}(t,t')=&&\langle X^\dag_j(t')X_i(t)\rangle.
\nonumber\\
=&&\prod_{q=1} \Bigg\{
e^{-\frac{\lambda^2_{iq}+\lambda^2_{jq}}{2}(2N_q+1)}
\text{exp}
\Big\{\lambda_{iq}\lambda_{jq}
\times\nonumber\\
&&\Big[
N_q e^{-i\omega_q(t-t')}
+(N_q+1)e^{i\omega_q(t-t')}
\Big]\Big\}\Bigg\},
\end{eqnarray}
where $N_q$ is the occupation number for the $q$th phonon mode.
$K^<_{ij}(t,t')$ can be decomposed as
\begin{eqnarray}\label{eq:shiftcorrelation2}
K^<_{ij}(t,t')
=&&\prod^M_{q=1}
\left[ \sum_{p_q}
L^{p_q}_{ij} e^{i p_q\omega_q(t-t')}
\right]
\nonumber\\
=&&\sum_{p_1 p_2\cdots p_M}
L^{p_1}_{ij}L^{p_2}_{ij}\cdots
L^{p_M}_{ij}e^{i\bm{p}^\text{T} \bm{\omega}(t-t')}
\nonumber\\
\equiv&&
\sum_{\bm{p}}L^{\bm{p}}_{ij}e^{i\bm{p}^\text{T} \bm{\omega}(t-t')},
\end{eqnarray}
where both $\bm{p}$ and $\bm{\omega}$
are row vectors,
$\bm{p}^\text{T}\bm{\omega}=\sum_q p_q\omega_q$.
And $L^{\bm{p}}_{ij}=L^{p_1}_{ij}L^{p_2}_{ij}\cdots
L^{p_M}_{ij}$, where
$L^{p_q}_{ij}$ is modified Bessel function
\begin{eqnarray}
L^{p_q}_{ij}=&&e^{-\frac{\lambda^2_{iq}+\lambda^2_{jq}}{2}(2N_q+1)}
e^{p_q\omega_q\beta/2}
\nonumber\\&&\times
I_{p_q}\left[2\lambda_{iq}\lambda_{jq}\sqrt{N_q(N_q+1)}\right],
\end{eqnarray}
and $I_{p_q}$ is the $p_q$th order Bessel function.

The greater projection of displacement correlation function is
$
K^>_{ij}(t,t')=\langle X_i(t)X^\dag_j(t')\rangle
=[K^<_{ij}(t,t')]^\dag.
$
It can be verified that $K^<(t,t')\simeq K^>(t,t')$ in the
high-temperature limit since $N_q\simeq N_q +1$ in this limit.
Actually, many works even adopt $K^<(t,t')\simeq K^>(t,t')$
as an approximation in low temperature, which is
regarded as neglecting the Fermi Sea
\cite{huanghartmut2008,PhysRevB.66.075303,
PhysRevB.67.165326,PhysRevB.40.11834,PhysRevB.71.165324}.

With $K^{</>}(t,t')$ known, the lesser (greater) self-energy
in energy space can be obtained from the Fourier transformation
of $\Sigma^{</>}_{I\alpha}(t,t')$,
\begin{eqnarray}
\bar{\Sigma}^<_{I\alpha,ij}(E)=&&\sum_{\bm{p}} L^{\bm{p}}_{ji}
\Sigma^{<}_{\alpha,ij}(E-\bm{p}^T\bm{\omega})-
\bar{\Sigma}^{<}_{0\alpha,ij}(E),
\nonumber\\
\bar{\Sigma}^>_{I\alpha,ij}(E)=&&\sum_{\bm{p}} L^{\bm{p}}_{ji}
\Sigma^{>}_{\alpha,ij}(E+\bm{p}^T\bm{\omega})-
\bar{\Sigma}^{>}_{0\alpha,ij}(E).
\end{eqnarray}

\subsection{Observable of interest}
The steady-state current through the junction can
be expressed in Meir-Wingreen formula~\cite{PhysRevB.50.5528}
\begin{equation}
I_\alpha=\frac{2e}{\hbar}\int \frac{dE}{2\pi}
T_\alpha(E)
\end{equation}
where the factor $2$ accounts for the spin degeneracy
and $T_\alpha$ is the transmission coefficient through
lead $\alpha$, which is expressed in terms of Green's
functions and self-energies within NEGF formalism
\begin{equation}\label{eq:trans}
T_\alpha(E)=\text{Tr}\left[
\Sigma^<_\alpha(E)G^>(E)-\Sigma^>_\alpha(E)G^<(E)
\right].
\end{equation}
As shown in Eq.(\ref{eq:green0}), lesser/greater projection of
Green's functions are
\[
G^{\lessgtr}_{ij}(t,t')=\bar{G}^{\lessgtr}_{ij}(t,t')
K^{\lessgtr}_{ij}(t,t').
\]
Utilizing the expansion of $K^{\lessgtr}_{ij}(t,t')$,
Fourier transformation of above equations gives
the Green's functions as
\begin{eqnarray}\label{eq:greenH}
G^>_{ij}(E)=&&\sum_{\bm{p}} L^{\bm{p}}_{ij}
\bar{G}^>_{ij}(E-\bm{p}^T\bm{\omega}),\nonumber\\
G^<_{ij}(E)=&&\sum_{\bm{p}} L^{\bm{p}}_{ij}
\bar{G}^<_{ij}(E+\bm{p}^T\bm{\omega}).
\end{eqnarray}
Hence, transmission coefficient and current can be
obtained through substituting Eq.(\ref{eq:greenH}) into
Eq.(\ref{eq:trans}).

In short, the computational procedures are summarized as
follows:
(1)~Firstly, bare Green's functions $\bar{G}^{r,\lessgtr}_0(E)$
in variational polaron picture are calculated; (2) Secondly,
Green's functions $\bar{G}^{r,\lessgtr}(E)$ at the second-order
with respect to $\bar{H}_I$ are evaluated; (3) Thirdly,
Green's functions in the $H$ picture are obtained
through Eqs.(\ref{eq:greenH}); (4) Finally, physical
quantities, such as density of states (DOS) and currents,
are obtained from the Green's functions and self-energies.

\section{Numerical examples}
\label{sec:numerical}
\begin{figure}[!htb]
  \includegraphics[width=0.48\textwidth]{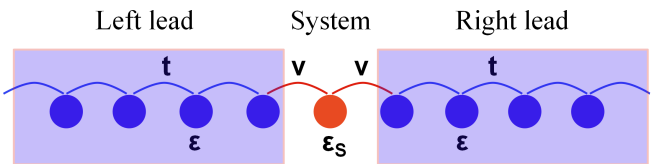}
  \caption{Illustration of a single site model coupled
  to two leads. It is assumed that the system coupled
  to the two lead equally, which is denoted by $v$. The
  onsite energy and hopping element of leads are
  $\epsilon$ and $t$, respectively.}
  \label{fig:linearchain}
\end{figure}

The variational polaron theory for dissipative quantum
transport is employed to study the transport through on a
monoatomic chain. The system consists of a single site.
The two leads are modeled by semi-infinite linear chains as
illustrated in Fig.~\ref{fig:linearchain}.
The onsite energy of the system is denoted by $\epsilon_S$.
The onsite energy and hopping element of leads are
$\epsilon$ and $t$, respectively. The system-lead coupling
constant is denoted by $V_\alpha$. In this work, it is
assumed that the system is coupled to the two leads equally,
i.e., $V_L=V_R=v$.
The line-width function of the semi-infinite tight-binding
lead can be evaluated analytically~\cite{scheer2010molecular}
\begin{equation}
\Gamma_\alpha(E)=V^2_\alpha
\frac{\sqrt{4t^2-(E-\mu_\alpha)^2}}{t^2},
\end{equation}
where $\mu_\alpha$ is the chemical potential of
lead $\alpha$.

\begin{figure}[htb!]
  \includegraphics[width=0.48\textwidth]{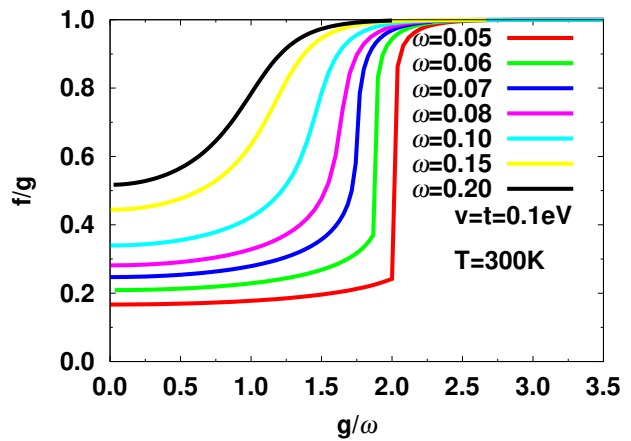}
  \vspace{-10pt}
  \caption{\label{fig:vptvalue}Optimal value
  $f$ of variational polaron transformation against
  different electron-phonon coupling constant $g$.
  Parameters are: The system-lead coupling
  and hopping in leads are $v=t=0.1$~eV;
  $T=300$~K.}
  \vspace{-10pt}
\end{figure}

In this work, the system-lead coupling is set to be equal to
the hopping in the leads, i.e., $v=t$. The onsite energy is set
to be $0$~eV and hopping is set as $t=0.1$~eV. And the Fermi
energy of the leads is $0$~eV, which corresponds to
half-filled on each site. Temperature is set as $T=300$~K.
In this work, only one phonon mode is considered to be coupled
to the system for simplification. The optimal value
$f$ of variational polaron transformation is searched for
different phonon frequency and electron-phonon coupling
strength. The phonon frequency is chosen in three different
regimes:
$\omega>t$, $\omega=t$ and $\omega<t$. The optimal value $f$
of variational polaron transformation for different
electron-phonon coupling constant $g$ is shown in
Fig.~\ref{fig:vptvalue}. As it is expected, different
electron-phonon coupling strength corresponds to different
optimal value of $f$ in the variational transformation. Besides,
$f$ increases monotonously with electron-phonon coupling
strength $g$. When $g\rightarrow 0$,
$f/g$ ratio approaches a small value which is affected
by the phonon energy. In the $g\rightarrow 0$ limit,
$f/g$ ratio is small, which is close to no polaron
transformation. While in the strong
electron-phonon coupling limit ($g\rightarrow\infty$),
$f/g$ ratio approaches to 1, which corresponds to full
polaron transformation. In the intermediate regime,
variational polaron transformation ends up with
$f/g<1$, which denotes the partial polaron
transformation.

\begin{figure}[htb!]
  \includegraphics[width=0.48\textwidth]{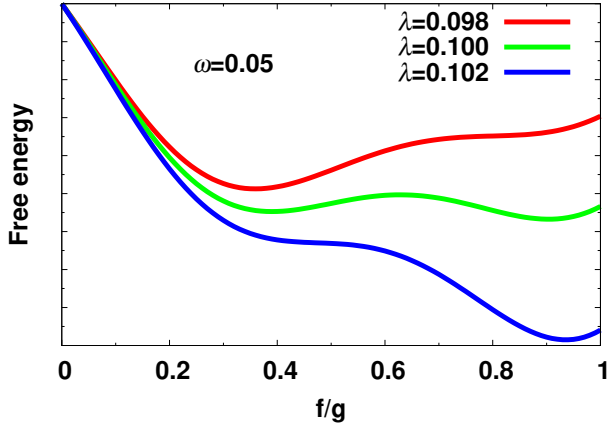}
  \caption{Free energy as function of variational
  transformation parameter $f$ for different electron-phonon
  coupling strength $g$.
  The phonon energy is $0.05$~eV. When $g$ is around
  $0.10\sim 0.102$~eV, different local minimum may become the
  global minimum, resulting in abrupt transition of $f$.}
  \label{fig:free}
\end{figure}

It is also noticed from Fig.~\ref{fig:vptvalue} that
there is a shape transition of $f$ in a certain regime of
$g$ if $\omega<t$.
Larger $\frac{t}{\omega}$ results in the steeper transition.
Finally, $f$ shows discontinuity against
$\frac{g}{\omega}$. This discontinuity appears because
different local minimum becomes the global
minimum at certain parameter regime. Taking
$\omega=0.05$~eV as example, the free energies against
$f/g$ ratio with electron-phonon coupling strength
around $0.1$~eV are plotted in Fig.~\ref{fig:free}.
As it shown by Fig.~\ref{fig:free}, two local minimums
start to emerge at $g\approx 0.1$~eV. The local
minimums of smaller and larger $f/g$ become the
global minimum for $g=0.098$~eV and $g=0.102$~eV,
respectively. The discontinuity of $f$ may cause unphysical
phase transition from delocalized to localized
regime~\cite{1.433182,1.4722336,PhysRevB.85.224301}.
For multiple
sites systems, the free energy can be more complex and more
local minimums may emerge, thus there is larger possibility
of discontinuity. However, even the discontinuity appears,
the variational transformation is still likely to work better
than the untransformed scheme or polaron transformation.
Moreover, the discontinuity of $f$ may be removed by employing
certain ansatz. For instance, a ground state ansatz has been
proposed recently to get rid of the discontinuity in the
variational polaron~\cite{PhysRevB.85.224301,PhysRevB.90.075110}.
Similar ansatz may be proposed to remove the
discontinuity problem in the quantum transport system, which
is the scope of further improvement of the method.

%
After the variational polaron transformation is found,
the quantum transport properties of the system, such as
DOS and current-voltage characteristics,
at different electron-phonon coupling regime can be studied
by the method developed in Sec.~\ref{sec3}.
Here, three sets of parameters are studied, ranging from
weak to strong electron-phonon coupling regime. The parameters
as well as optimal factor of variational polaron transformation
are summarized in Table~\ref{tab:models}.
Model A and C are in the weak and strong coupling regime,
respectively, and Model B is in the moderate coupling
regime. The optimal values of variational polaron
transformation are $0.17$, $0.54$ and $0.90$.
Here, the $g/\omega$ ratio is set as a constant for the
three models. As shown by Fig.~\ref{fig:vptvalue},
large $g/\omega$ ratio ($g/\omega\gg 1$) makes the system
in polaron regime while small $g/\omega$ ratio
($g/\omega\ll 1$) makes the system in weak or moderate
coupling regime. Hence, $g/\omega=0.8$ is chosen
to make sure that the system varies from the weak to strong
coupling by tuning the electron-phonon coupling constant.

\begin{table}[htb!]
\caption{\label{tab:models}
Parameters of the Models.
}
\begin{ruledtabular}
\begin{tabular}{cccc}
\textrm{Model}&
\textrm{$\omega$~(eV)}  &
\textrm{$g$~(eV)}  &
\textrm{$f/g$}  \\
\colrule
 A & $0.05$ & $0.04$ & $0.17$ \\
 B & $0.15$ & $0.12$ & $0.54$ \\
 C & $0.40$ & $0.32$ & $0.90$
\end{tabular}
\end{ruledtabular}
\end{table}

\begin{figure}[htb!]
  \includegraphics[width=0.48\textwidth]{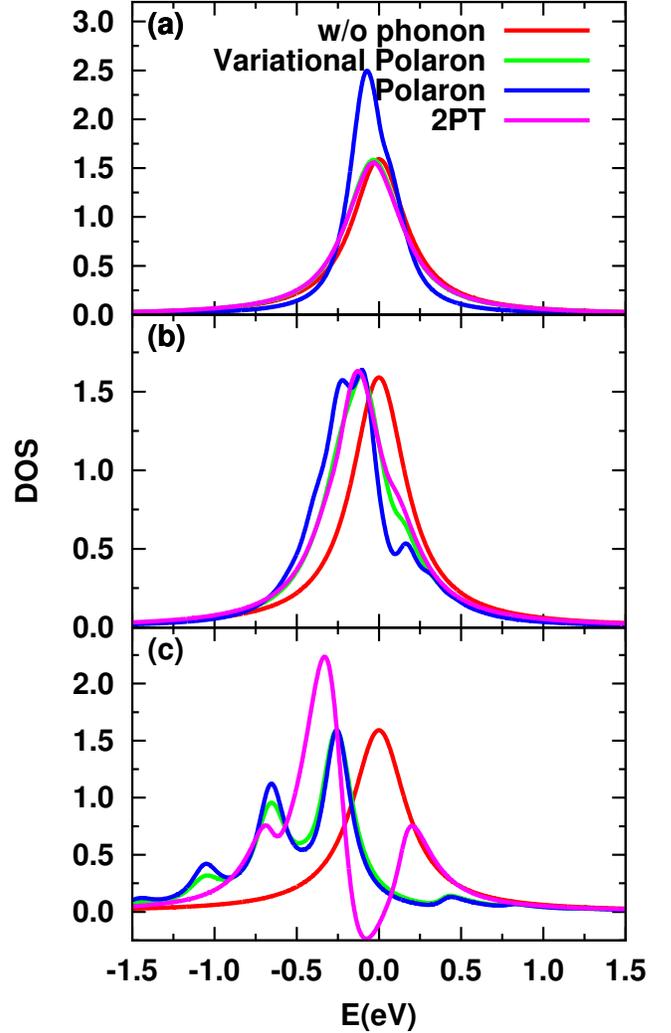}
  \caption{\label{fig:DOS}
  DOS of the system with different electron-phonon
  coupling constant:
  (a) $\omega=0.05$~eV and $g=0.04$~eV;
  The electron-phonon interaction is in the weak coupling regime.
  DOS of variational polaron is same as that of 2PT, while
  polaron theory overestimates the phonon side-bands.
  (b) $\omega=0.15$~eV and $g=0.12$~eV;
  In this moderate electron-phonon coupling regime, 2PT
  starts to fail and polaron still overestimates the phonon
  side-bands.
  (c) $\omega=0.40$~eV and $g=0.32$~eV.
  The system is in the strong coupling regime. Variational
  polaron theory gives similar results as polaron theory does.
  While 2PT is not applicable to this regime, it even
  gives unphysical DOS.
  See text for other parameters.
  }
\end{figure}

In presence of electron-phonon coupling,
electronic structure of the system is affected
at different levels depending on the electron-phonon
coupling strength. Consequently, DOS of the system is
affected, such effects include
energy shift of peak, broadening of the spectral and
emergence of phonon side-bands.
Fig.~\ref{fig:DOS} plots the DOS of the system
with different electron-phonon coupling strength.

(a) In model A, the electron-phonon coupling strength
is small. Phonon has limited effects on the electronic
structure of the system. The phonon-induced energy shift
is about  $50$~meV.
The DOS of variational polaron transformation is
close to that of 2PT. Moreover, the DOS of both
variational polaron theory and 2PT is almost the 
same as that of non-interacting system except
the phonon-induced shift of the peak. No obvious
phonon side-bands are observed in the DOS of
variational polaron theory and 2PT.
However, polaron theory overestimates the energy shift
and phonon side-bands. Because polaron theory overestimates
the phonon dressed system-lead coupling, resulting in
stronger localization effect. Consequently, main peak
of DOS is sharped and phonon side-band appears as shown in
Fig.~\ref{fig:DOS}(a).

(b) In the moderate coupling regime, phonon has
larger effect on the electronic structure of the system
and phonon side-bands start to appear due to the increased
phonon-induced electron localization.
In this coupling regime, the DOS of variational
polaron theory starts to deviate from that of 2PT as shown
by Fig.~\ref{fig:DOS}(b).
Due to the principle of 2PT,
only two phonon side-bands can be observed in the DOS.
However, higher order corrections become more and more
significant in the moderate coupling regime. As a
result, 2PT starts to fail to describe the renormalization
effect induced by phonon.
Besides the phonon side-bands, the energy shift due to
electron-phonon interaction is underestimated slightly by 2PT.
In contrast, higher order phonon side-bands can be evaluated
by both polaron theory and variational polaron theory.
But, polaron theory still overestimates
the phonon dressed system-lead coupling compared
with variational polaron theory. Consequently,
the phonon side-bands are much more significant
in the DOS. While, variational polaron theory
neither underestimates nor overestimates the
renormalization effect. The DOS of variational polaron
lies between those of 2PT and polaron.

(c) In the strong coupling limit, phonon induces
pronounced side-bands effect and polaron is formed.
Consequently, the variational polaron transformation
approaches the polaron theory.
As shown by Fig.~\ref{fig:DOS}(c), DOS of variational
polaron theory is almost the same as that of polaron theory
since the optimal value of $f/g$ is $0.9$, which is close
to $1$, i.e., variational polaron theory approaches the
polaron theory. In the contrast, 2PT greatly underestimates
the phonon side-bands and overestimates the phonon-induced
energy shift. The pronounced phonon side-bands indicate
that electron is localized by the phonon. The phonon
side-bands is also responsible for the phonon blockade
in the current-voltage characteristics, which is represented
by steps in the current-voltage characteristics or peaks in
the differential conductance~\cite{PhysRevB.71.165324,
PhysRevB.66.075303,PhysRevB.67.165326,nitzan2006prb}.

\section{Summary}
\label{sec:summary}
In this work, a dissipative quantum transport theory
with electron-phonon coupling at arbitrary parameter regime is
developed by employing the variational polaron transformation.
Variational polaron transformation provides an optimal set
of parameters for the unitary transformation,
which is determined by the minimization of the
Feynman-Bogoliubov upper bound of free energy.
After the minimization, variational polaron transformation
is able to end up with an optimal mean-field
Hamiltonian and  a small perturbation.
The mean-field part Hamiltonian, $\bar{H}_0$, contains
the phonon-induced energy renormalization and phonon dressed
system-lead coupling. Thanks to the variational polaron
transformation, the many-body effect
of electron-phonon interaction is transformed into the
small interacting part of the Hamiltonian ($\bar{H}_I$),
which validates the 2PT treatment.
Upon the variational transformation, a quantum transport
theory with electron-phonon interaction at arbitrary
parameter regime is established within NEGF formalism.
Following the quantum transport theory of non-interaction
system, the mean-field part is treated exactly. $\bar{H}_I$
is taken into account through the self-energies which
are in second-order of $\bar{H}_I$

Numerical examples on a monoatomic chain demonstrate the
validity of the variational transformation.
The optimal value of variational transformation
increases monotonously with electron-phonon coupling
strength, which connects
the 2PT in weak coupling regime
and polaron theory in strong coupling limit.
Comparison of DOS in different parameter regimes
indicates the applicability of variational polaron
transformation: (a) In weak coupling limit,
polaron theory overestimates the phonon-induced energy
shift and side-bands; (b) In the strong coupling limit,
2PT underestimates the renormalization
effect induced by phonon and even gives unphysical DOS;
(c) Variational polaron transformation naturally connects
the 2PT in the weak coupling limit and polaron theory in
the strong coupling limit by making use of the optimal
transformation.
More numerical benchmark comparison
with other numerically exact methods,
such as QUAPI and MCTDH,
will be considered in the future works. Moreover,
the sudden change of the optimal value of variational
transformation may be solved by employing certain
ansatz~\cite{PhysRevB.85.224301,PhysRevB.90.075110},
which is also the scope of future improvement of the method.

\begin{acknowledgments}
Y.~Zhang thanks Jianshu Cao, Weitao Yang, E.~K.~U. Gross
and Garnet Chan for helpful
discussions. The support from the
Hong Kong Research Grant
Council (Contract Nos. HKU7009/09P, 7009/12P, 7007/11P, and
HKUST9/CRF/11G (G.H.~Chen)), the University Grant Council
(Contract No.~AoE/P-04/08 (G.H.~Chen)), National Natural Science
Foundation of China (NSFC 21322306 (C.Y.~Yam), NSFC
21273186 (G.H.~Chen, C.Y.~Yam)), and National Basic Research Program
of China (2014CB921402 (C.Y.~Yam))
is gratefully acknowledged.
\end{acknowledgments}

\bibliography{ref}
\end{document}